\documentclass[useAMS,usenatbib]{heliumpaper}

\usepackage{graphicx}
\usepackage{grffile}
\usepackage{epsf}
\usepackage{amsmath}
\usepackage{amssymb}
\usepackage{bm}
\usepackage{enumitem}
\usepackage[normalem]{ulem}

\usepackage{hyperref}      
\usepackage[all]{hypcap}   

\setlist[enumerate,1]{leftmargin=*}

\newcommand{\hkpc}{h^{-1}{\rm kpc}}

\newcommand{\kms}{{\rm km}\,{\rm s}^{-1}}

\newcommand{\TD}{{\sc Technicolor Dawn}~}
\newcommand{\CIV}{\hbox{C\,{\sc iv}}~}
\newcommand{\CIVb}{\hbox{C\,{\sc iv}}}
\newcommand{\CII}{\hbox{C\,{\sc ii}}}
\newcommand{\CIII}{\hbox{C\,{\sc iii}}~}
\newcommand{\MgII}{{\hbox{Mg\,{\sc ii}}}}
\newcommand{\SiII}{{\hbox{Si\,{\sc ii}}}}

\newcommand{\SiIV}{\hbox{Si\,{\sc iv}}~}
\newcommand{\SiIVb}{\hbox{Si\,{\sc iv}}}
\newcommand{\OI}{{\hbox{O\,{\sc i}}~}}
\newcommand{\HI}{{\hbox{H\,{\sc i}}~}}

\newcommand{\HeII}{{\hbox{He\,{\sc ii}}~}}
\newcommand{\HeIII}{{\hbox{He\,{\sc iii}}~}}
\newcommand{\HeIIb}{{\hbox{He\,{\sc ii}}}}

\graphicspath{{./}{Plots}}

\begin{document}

\title[3D UVB model with Helium {\textit{\sc ii}}]{Impact of Helium {\sc ii} resonant absorption on the UVB modeled in three dimensions}

\author[E. Huscher et al.]{
\parbox[t]{\textwidth}{\vspace{-1cm}
Ezra Huscher$^{1,2}$, Kristian Finlator$^{1,2}$, Samir Ku\v{s}mi\'{c}$^{1}$, Maya Steen$^{1}$
}\\\\
$^1$Department of Astronomy, New Mexico State University, Las Cruces, NM, USA\\
$^2$Cosmic Dawn Center (DAWN), Niels Bohr Institute, University of Copenhagen, Denmark
}
\maketitle

\pubyear{2024}

\maketitle

\label{firstpage}

\begin{abstract}

We implement a treatment of Helium {\sc ii} absorption and re-emission into the \TD cosmological simulations to study its impact on the metagalactic ultraviolet background (UVB) in three dimensions. By comparing simulations with and without He {\sc ii} reprocessing, we show that it weakens the mean UVB by $\sim$3 dex from $z = 10$ to $z = 5$ between 3.5 and 4 Ryd, where the He {\sc ii} Lyman-series resonance occurs. In overdense regions, the overall UVB amplitude is higher and the impact of \HeII reprocessing is weaker, qualitatively indicating an early start to \HeII reionization near galaxies. Comparing our simulations to two popular one-dimensional UV models, we find good agreement up to 3 Ryd at $z = 5$. At higher energies, our simulation shows significantly greater He {\sc ii} absorption because it accounts for He {\sc ii} arising both in diffuse regions and in Lyman limit systems. By contrast, the comparison models account only for He {\sc ii} in Lyman limit systems, which are subdominant prior to the completion of He {\sc ii} reionization. The H {\sc i} and He {\sc ii} reionization histories are nearly unaffected by He {\sc ii} reprocessing although the cosmic star formation rate density is altered by up to $4\%$. The cosmic mass density of C {\sc iv} is reduced by $\sim$2 dex when He {\sc ii} is accounted for while Si {\sc iv}, C {\sc ii}, Mg {\sc ii}, Si {\sc ii}, and O {\sc i} are unaffected.

\noindent
\end{abstract}

\section{Introduction}
\label{sec:intro}

The Universe is filled with radiation accumulated since the Big Bang, providing precious hints to its history and evolution. Certain sources, such as massive stars and supermassive black holes, generate radiation in the ultraviolet which can be filtered, re-processed, and redshifted over time. This ultraviolet background (UVB) impacts galaxy formation and drives reionization, making it crucial to understanding how the Universe evolved to its present state.

Early groundwork for observationally inferring the UVB was laid by \cite{shapiro87}, who showed that galaxies are the primary ionizing sources at high redshift with quasars becoming more important in later times. \cite{hm96} developed a technique for integrating the cosmological radiative transfer equation that accounts for the emission from different sources and absorption from different sinks so as to predict, in an empirically-driven way, the evolution of the UVB with redshift. A number of synthesis UVB treatments followed including \cite{fardal98}, \cite{mh09}, \cite{hm12} (HM12 hereafter), \cite{khaire13}, and \cite{fg20} (FG20 hereafter).  These one-dimensional models account for a wide range of physical processes affecting the IGM's thermal and ionization state, they readily offer high spectral resolution, and they are well-calibrated by design because they treat the emissivity and absorber abundances as observational inputs. Their outputs are routinely used to compute the ionization states of the intergalactic medium (IGM) and circumgalactic medium (CGM) and for modeling the cooling of gas in cosmological hydrodynamic simulations \citep{somerville15}.

Despite their efficiency and realism, however, synthesis UVB models address a number of processes incompletely or not at all. For example, the observational calibration to the observed Lyman-$\alpha$ forest column density distribution neglects the feedback effects between UVB sources and sinks. The assumption of spatial homogeneity prevents them from accounting for source-sink clustering, which can suppress the volume-averaged UVB. Likewise, they do not capture spatial fluctuations in the UVB's amplitude or slope. Such fluctuations are inevitable prior to and in the immediate aftermath of reionization and can influence the abundance of observable metal absorbers~\citep{finlator2015}.

One sink ion whose treatment remains incomplete in most one-dimensional UVB models is singly-ionized helium (He {\sc ii}). Four relevant processes can ensue when ultraviolet flux encounters this ion:
\begin{enumerate}
    \item He {\sc ii} Lyman-series transitions absorb UV flux between 3--4 Ryd
    \item A portion of the flux absorbed from the first set of processes is released as He {\sc ii} Ly-$\alpha$ re-emission at 3 Ryd
    \item He {\sc ii} Balmer continuum recombination ($n=\infty\rightarrow2$) results in ionizing flux at $\sim$1 Ryd
    \item He {\sc ii} recombinations can additionally lead to Ly-$\alpha$ emissions at 3 Ryd
\end{enumerate}

He {\sc ii} provides an important window into the early Universe because it is sensitive to high-energy flux at an epoch when quasars were not yet abundant, allowing study of the other primary sources capable of producing UV flux at energies $>$3.5 Ryd: first-generation, massive, low-metallicity stars. Its fine structure is sixteen times wider and Doppler widths fifty percent smaller than the corresponding structure in hydrogen \citep{syed12} which makes it a potentially valuable complement to H$\alpha$ emission in studying the cosmic star formation history, although the He Lyman-series is overwhelmed by the \HI Lyman-$\alpha$ forest at $z>5$.

As shown by~\citet{mh09}, in models where \HeII is accounted for, the UVB intensity peaks at frequencies just above each transition's resonant frequency, vanishes at resonance, then rises with decreasing energy until the next resonance is encountered, leading to a distinct ``sawtooth" structure in UVB. The cosmic abundance of He {\sc ii} affects the efficiency of this process, as photons continue to scatter until they redshift out of resonance.~\citet{mh09} suggested that this reprocessing could, in turn, affect the abundance of metal ions that are sensitive to flux with energies between 3--4 Ryd. 

Modeling He {\sc ii} reprocessing is challenging because it resides predominantly in diffuse gas prior to the completion of He {\sc ii} reionization at $z\approx2.8$~\citep{worseck2019}. By contrast, \HI is dominated by diffuse gas prior to the completion of \HI reionization at $z\approx5.3$~\citep{bosman22}, and then by Lyman limit systems afterwards. Modeling these effects requires separate treatments for the ionization states of diffuse and dense gas, and separate treatments for \HI and He {\sc ii} reionization histories. However, most synthesis UVB models make the simplifying assumption that the cosmic opacity including its He {\sc ii} contribution is confined to \HI Lyman limit systems at all times. This can lead to an underestimate of the opacity from diffuse gas prior to reionization.

This issue was partially addressed by \citealt{puchwein19}, who combined three-dimensional simulations with an updated version of the HM12 model. This ``best-of-both-worlds" approach enabled them to model the opacity from diffuse gas separately from the Lyman limit systems, adding significant realism to the predicted UVB. They distilled their results into improved volume-averaged photoionization and photoheating rates for uptake by cosmological simulations. Unfortunately, their predictions were not published in a way that enables readers to assess the impact on metal ion absorbers.

More generally, \textit{the effect of He {\sc \textit{ii}} opacity on metal ion absorbers in the CGM has not been studied in cosmological simulations with a live, self-consistent UVB.} In exchange for unavoidable limitations in dynamic range and spectral resolution, three-dimensional simulations provide increased realism by allowing localized fluctuations in the UVB's intensity and spectral slope, all of which impact ion abundances in the CGM. 

The abundances of \CIV and \SiIV are particularly useful observationally. Both have transitions that can be detected in high-redshift CGM which allows us to examine the CGM's density, temperature and metallicity \citep{mallik23}. However, the kinship between the two ions diverges when their abundances are calculated in cosmological models. \SiIV matches well in {\sc Eagle} \citep{schaye15} and {\sc Illustris} \citep{nelson15} simulations, but \CIV does not, being underproduced by 20x in {\sc Eagle} \citep{rahmati16} and 10x in \TD \citep{finlator20}. It is not certain why \CIV has been more challenging to reproduce, but it is conceivably some combination of two scenarios: 1) either simulations are not accurately reproducing the ionization sources which would transform more of the existent carbon into the triply-ionized state, or 2) there is more carbon in space than we expect due to incomplete knowledge of carbon production, such as the difficult-to-observe thermal pulsing activity in AGB stars. 


Considering the first possibility, \citealt{diaz20} presented evidence that strong \CIV absorbers are found more often around faint galaxies rather than bright ones because faint galaxies eject more metals into their CGM via outflows. However, if strong \CIV absorbers are indeed associated primarily with faint galaxies, then high-resolution cosmological simulations coupled with accurate UVB models like \TD should not underproduce them. If anything, the tendency for cosmological simulations to underproduce both the strong \CIV absorbers and the \CIV mass density points to an association with rare massive systems which are systematically underproduced by simulations with limited cosmological volumes~\citep{springel03b}.


Approaching this from another view, simulations can readily produce different column densities of ions such as \CIV by adjusting parametrized inputs to the feedback model including the ionizing escape fraction and feedback efficiency; by adjusting the UVB amplitude; or by accounting for small-scale UVB fluctuations \citep{opp09,finlator2016,keating16,finlator20}.Calibrating these choices to other expected markers, such as the strength of the Lyman alpha forest transmission and the galaxy stellar mass function, is the foundational approach of simulation design. After aligning these metrics to observational values, \CIV is the ion that is usually underproduced by up to 20x in simulations when compared to observations while \SiIV and other ions match well (\citealt{rahmati16}; see, however,~\citealt{garcia2017}). Studies have repeatedly shown that \CIV can be only be matched to observations by losing alignment to these other observables, and therefore it has become a frequent loose end in high-redshift theory.

While adjusting feedback models should continue to be considered, we take the complementary approach of improving the model's physical realism, enabling us to characterize the discrepancy more completely. Previous works have shown that \HeII re-processing modulates the UVB in the energy range relevant for \CIII ionization~\citep{puchwein19} and that \CIV is significantly enhanced by UVB inhomogeneity \citep{opp09,finlator2016}. Not until this work, however, has UVB inhomogeneity and \HeII reprocessing been modeled together to explore the effect on \CIV abundance.

\begin{figure*}
    \includegraphics[width=0.97\textwidth]{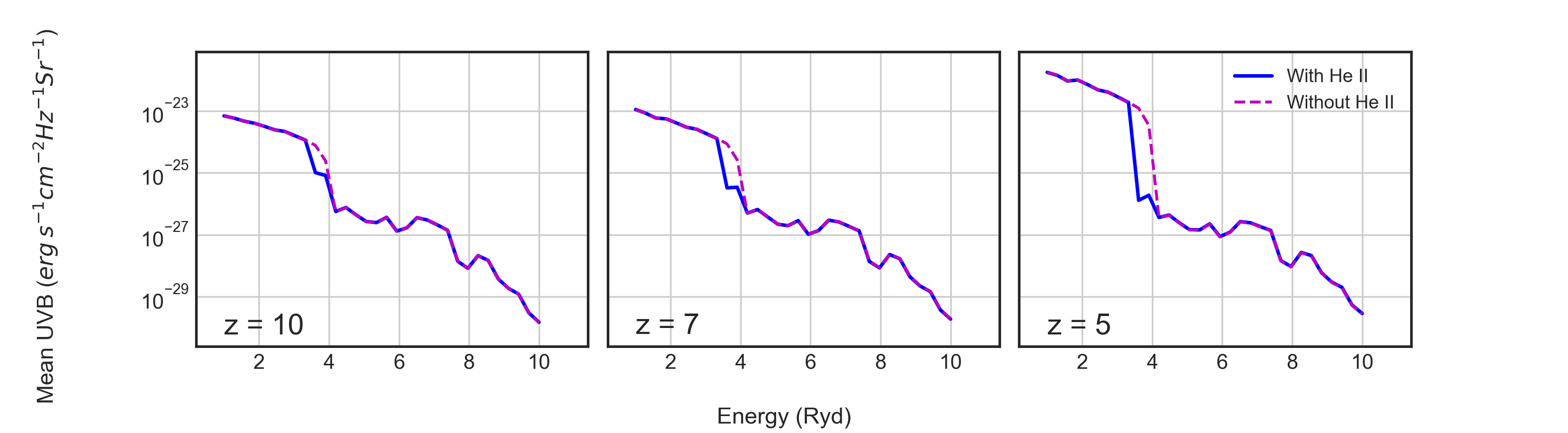}
    \caption{ Mean specific radiative UVB intensity with and without the resonant Helium {\sc ii} Lyman-series absorption. The absorption of the UVB at $\sim 3.5$ Ryd from this process begins in earnest at z=10. The UVB intensity is reduced by 1 dex at z=10, 2 dex by z=7, and 3 dex by z=5. }
    \label{fig:meanUVB}
\end{figure*}

We have accordingly implemented He {\sc ii} reprocessing into the \TD radiation transport module, enabling us to model directly its impact on the abundance of C {\sc iv}, Si {\sc iv}, C {\sc ii}, Mg {\sc ii}, Si {\sc ii}, and O {\sc i}. It is notable that the He {\sc ii} ionization potential (54.4 eV) lies squarely between 47.9 and 64.5 eV, the ionization energies of C {\sc iii} and \CIV respectively. Indeed, we will show that the abundance of \CIV is by far the most sensitive to \HeII and that \HeII opacity significantly suppresses its abundance. In Section \ref{sec:simulation}, we present recent improvements to the \TD simulation. In Section \ref{sec:methods}, we present the methods used. In Section \ref{sec:results}, we discuss our results. We conclude and comment on future directions in Section~\ref{sec:summary}.

\section{Simulation}
\label{sec:simulation}

 The simulation framework that we use, \TD ({\sc TD}; \citealt{finlator2018}), combines a mature model for galaxy evolution in a cosmological context with an on-the-fly radiation transport solver that is optimized to model the UVB with unusually high spectral resolution. This accounts naturally for UVB spatial fluctuations as well as nonlinear interactions between the radiation field and galaxy formation. {\sc TD} was developed by modifying Gadget-3, a well-established smoothed particle hydrodynamics framework \citep{springel05}. It incorporates a number of upgrades with respect to our previous work~\citep{finlator20}, which we outline here. The treatment for the QSO contribution to the UVB is unchanged with respect to~\citet{finlator20}

\begin{figure}
    \includegraphics[width=0.49\textwidth]{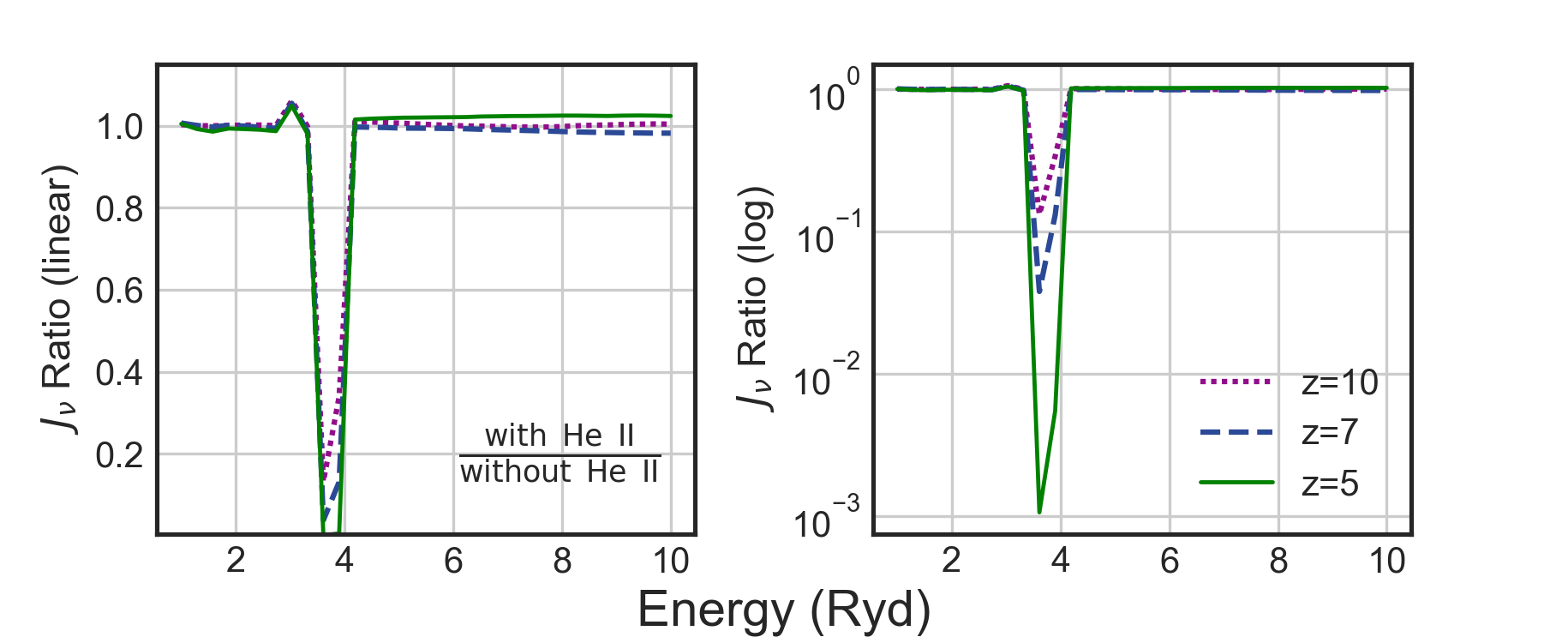}
    \caption{ The ratio of the TD model with to without He {\sc ii} reprocessing at each of the 32 UVB bins with a linear y-axis (left) and logarithmic y-axis (right). The linear panel accentuates the bump from \HeII Ly$\alpha$ re-emission at $\sim 3$ Ryd.}
    \label{fig:ratioUVB}
\end{figure}

\subsection{Higher UVB Spectral Resolution}\label{ssec:resolution}
Whereas in previous work we discretized the UVB into 24 energy bins evenly spanning energies from 1--10 Ryd, we now increase this to 32 bins. The increased resolution enables our model to capture more structure in the UVB's shape, including the first indications of the sawtooth feature of the \HeII Lyman-series as described by \citealt{mh09}.

\noindent
\subsection{Improved emissivity from binary stellar evolution}
\label{ssec:emissivity}

\noindent
Our new simulations assume that star-forming gas radiates with a metallicity-dependent emissivity that is derived from BPASS version 2.2.1 \citep{stanwayeld18}. This represents a change with respect to our previous simulations, which adopted emissivities from {\sc Yggdrasil}~\citep{zackrisson11}. BPASS treats a number of physical effects associated with binary stellar evolution. Broadly, the net result is that young, low-metallicity stellar populations are brighter and show a harder spectral index in the far-ultraviolet; that is, in the regime relevant to reionization. These changes accelerate reionization, heat the IGM, and tilt the ionization balance in the CGM toward high-ionization ions such as \CIV \citep{rosdahl18, doughty21}.

\noindent
\subsection{Emissivity and Ionizing Escape}\label{ssec:escapefraction}
\noindent
As in our previous work, we derive the simulated emissivity field from star-forming gas particles by assuming that each particle has been forming stars at its current rate for 100 Myr. The fraction of ionizing flux that escapes from stellar photospheres $f_\mathrm{esc}$ cannot be modeled accurately in cosmological simulations owing to dynamic range limitations \citep{finlator20}, yet it remains a critical parameter to accurately model the large-scale ionizing radiation from galaxies \citep{puchwein19}. While a constant (and energy-independent) $f_\mathrm{esc}$ can reconcile galaxy observations with the redshift at which reionization completes~\citep{robertson15}, simultaneously matching observations of the post-reionization Lyman--$\alpha$ forest requires a redshift-dependent $f_\mathrm{esc}$ that is usually treated as a free parameter~\citep{hm12, khaire16, kuhlen12, doussot17}. 

For our purposes, $f_\mathrm{esc}$ means the fraction of flux that escapes from stellar photospheres out to distances comparable to the resolution of our radiation transport solver, or 187.5 comoving $\hkpc$. This is somewhat larger than the $\sim10$ proper kpc virial radius of the halos that likely dominated reionization, so it should be less than what is obtained from efforts that only considered escape out to the virial radius \citep{ma98, kostyuk23}. We follow the convention of assuming that galaxies resemble ionization-bounded clouds such that $f_\mathrm{esc}$ is energy-independent, and that it varies only with redshift $z$ as suggested by SPHINX \citep{rosdahl18} and THESAN \citep{kannan21} simulation results. We tune this dependence to match observations of the IGM and of the history of reionization. Adopting the metallicity-dependent BPASS emissivities requires that we re-scale $f_\mathrm{esc}$ with respect to our previous work in order to retain agreement with observations:

\begin{equation}\label{eqn:fesc}
f_\mathrm{esc}(z) = 0.146 \left(\frac{1+z}{6}\right),
\end{equation}
\noindent
and we further require that $f_\mathrm{esc}$ cannot exceed 0.18. \citet{bouwens15} and \citet{mitra15} show that this value is sufficient to achieve H {\sc i} reionization by z $\sim 6$; our results are similar.

\vspace{3mm}

\subsection{He {\sc ii} Re-Processing}
\label{ssec:reprocessing}

\noindent
As our simulations model the UVB with 32 independent frequency bins, it is possible for us to spectrally resolve the He {\sc ii} reprocessing \citep{haardt12}. We will show that the effects are overall modest except for their impact on the abundance of metal ions that are sensitive to energies in the range 3--4 Ryd, but incorporating them is nonetheless an improvement in realism and, as far as we are aware, a first for cosmological simulations. We account for He {\sc ii} Lyman-series absorption, re-emission of He {\sc ii} Lyman-$\alpha$ flux at 3 Ryd following He {\sc ii} Lyman-series absorptions, and He {\sc ii} Balmer continuum emission near 1 Ryd. Our approach follows previous work~\citep{fg09, mh09, hm12}, with slight modifications to account for the fact that our energy bins are much larger than the typical resonant line profile. All diffuse emissivities are added back into the gridded emissivity field without modifying the assumed Eddington tensors. This effectively assumes that flux associated with the diffuse field flows in the same direction as flux from discrete sources.

\subsubsection{Lyman-$\alpha$ Absorption and Emission}
\label{sec:lymanalpha}

\noindent
If $J_\nu$ represents the number density of photons per unit frequency at frequency $\nu$, then the rate of change of $J_\nu$ owing to  absorption by a resonant transition with oscillator strength $f$ occurring in ions with number density $n$ is

\begin{equation}\label{eqn:abs1}
\frac{d J_\nu}{dt} = -c n \frac{\pi e^2 f}{m_e c} \phi(\nu) J_\nu
\end{equation}
where $\phi(\nu)$ is the line profile. If this absorption is modeled in an energy bin spanning width $\nu_1$ to $\nu_2$ that is much broader than the resonant line profile, then the change in number density of photons in that energy bin $J \equiv \int_{\nu_1}^{\nu_2}(J_\nu)$ is derived by averaging over many small energy intervals within that bin. The result is
\begin{equation}\label{eqn:abs2}
\frac{d J}{dt} = -c n J\frac{\pi e^2 f}{m_e c (\nu_2 - \nu_1)}
\end{equation}
Intuitively, this is equivalent to a scenario in which the resonant line profile $\phi(\nu)$ is a tophat of unit area spanning the energy bin that contains it. The actual absorption coefficients in each energy bin are given by the fraction on the right-hand side of Equation~\ref{eqn:abs2}. They depend on the spectral resolution and are thus computed at runtime. Oscillator strengths for He {\sc ii} Lyman-series transitions are taken from Table 5 of~\citet{verner04}, and we include transitions up to Lyman-12. If two or more transitions contribute opacity within a single energy bin, then their oscillator strengths are summed.

Soon after a He {\sc ii} ion has absorbed a photon via a Lyman-series transition, the excited electron cascades back down to the ground state. Some of these cascades result in the emission of a He {\sc ii} Lyman-$\alpha$ photon at 3 Ryd. We account for this following MH09 and HM12. The fraction $f_n$ of He {\sc ii} Lyman-series absorptions to energy level $n$ that results in emission of a He {\sc ii} Lyman-$\alpha$ is $f_n$ is taken from \citet{pritchard06}. We add this emissivity back into the radiation transport solver as a diffuse, inhomogeneous source. Following HM12, we assume no local absorption of \HeII Lyman-$\alpha$ photons, equivalent to assuming that they scatter resonantly into the wings of the absorption profile and then escape, subject from then on only to photoelectric absorption. 

\subsubsection{Balmer Continuum Emission}
\label{sec:balmer}

\noindent
When an electron recombines directly from the continuum to the $n=2$ energy state about a He nucleus, it emits a photon with energy near 1 Ryd. The emission line profile is temperature-dependent, and is broader at higher temperatures. We pre-compute this emission profile for our energy resolution following Equations E3 and E4 of \citet{fg09} at a range of temperatures spanning 1000--40,000K, enabling the simulation to interpolate the profile to each particle's temperature. We adopt the Balmer continuum cross section from the NORAD database\footnote{\url{https://norad.astronomy.osu.edu/}} \citep{nahar10}. This recombination emissivity is added back into the radiation transport solver as a diffuse, inhomogeneous source.

\section{Methods} \label{sec:methods}

\subsection{Simulation run "pl9n384" details} \label{sec:simulation_details}

Our simulations span a cubical volume of side length $9h^{-1}$ Mpc. Dark matter and baryons are modeled using $2\times384^3$ particles, while the radiation field is discretized spatially into a uniform grid with $48^{3}$ voxels and spectrally into 32 frequency bins uniformly spanning the range 1--10 Ryd. We adopt the following cosmological parameters: $(\Omega_M, \Omega_\Lambda, \Omega_b, H_0, X_H) = (0.3089, 0.6911, 0.0486, 67.74, 0.751)$.

We run two ``pl9n384'' simulations of identical resolution and volume to examine the simulated differences when He {\sc ii} absorption and re-emission is included versus when it is omitted. We then compare the results to predictions from HM12 and FG20. For details concerning the calculations involved in the TD simulation output, such as averaging the UV photon density over all the voxels as in Figure \ref{fig:compare_models}, we refer the reader to \citet{finlator2018}.

\subsection{Simulated Absorbers} \label{ssec:rayCasting}

We extract simulated absorber catalogs using a ray-casting approach as described in~Section 3.7 of \citet{finlator2018}. Sightlines span a velocity width of $10^7~\kms$ using $2~\kms$ pixels. Instrumental broadening is modeled by convolving with a Gaussian profile whose full-width at half-maximum is $10~\kms$. Uncertainty is modeled using an adopted signal-to-noise of 50 per pixel. Simulated absorbers are identified as regions where the flux summed over at least three consecutive pixels drops at least $5\sigma$ below the continuum. Equivalent widths are computed directly from the spectra while column densities are inferred via the Apparent Optical Depth method~\citep{savage1991}. 

\noindent

\begin{figure}
    \includegraphics[width=0.49\textwidth]{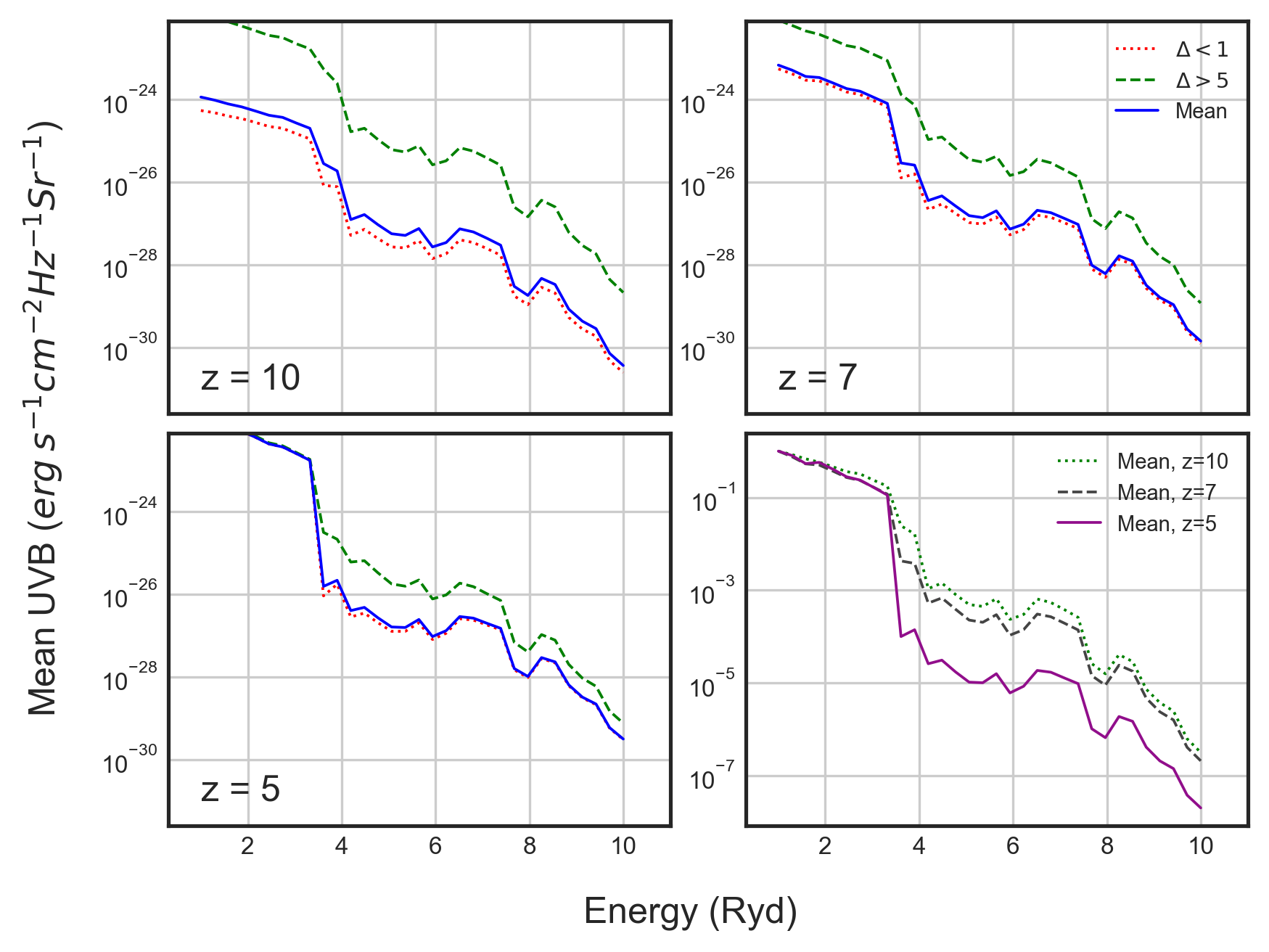}
    \caption{ The TD model with \HeII absorption added, divided into voids and overdense regions, at z=10, 7, and 5. The overdense regions show nearly 2 dex greater UVB than the mean. Underdense regions are only slightly lower than the mean, mostly at lower energies. The lower right panel compares the mean UVB normalized to the Lyman limit at each redshift. }
    \label{fig:overdensity}
\end{figure}

\begin{figure*}
    \includegraphics[width=0.97\textwidth]{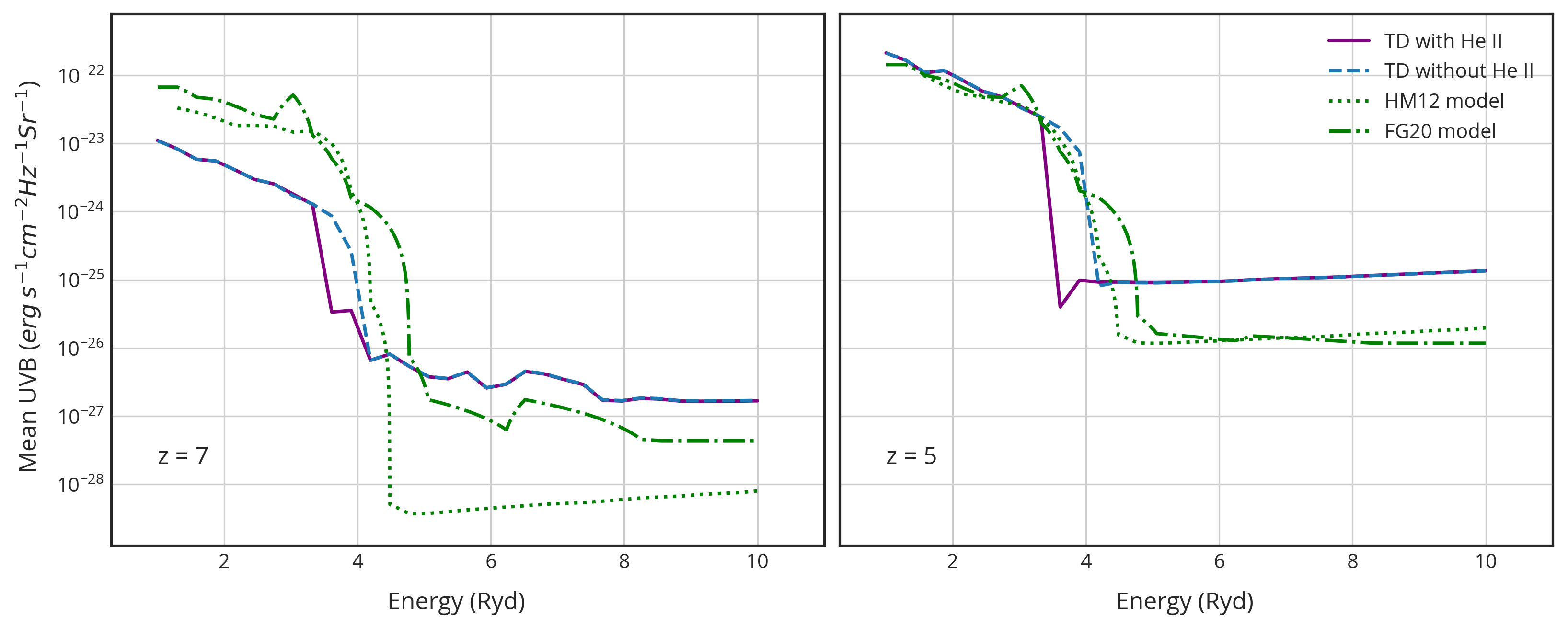}
    \caption{Comparing the volume-averaged \TD UVB at two redshifts with and without \HeII absorption and re-emission with the HM12~\citep[][dotted green]{hm12} and FG20~\citep[][dot-dashed green]{fg20}. The TD models include UV contributions from both galaxies and quasars because both are included in HM12 and FG20, but elsewhere in this paper quasar UV is omitted. Both HM12 and FG20 are interpolated over energy ranges where predictions were unavailable.
    Neither model accounts accurately for opacity owing to diffuse gas, which leads to their increased UVB in the lower energies at $z=7$. This difference disappears for energies below 3.5 Ryd by $z=5$ but lingers at higher energies.}
    \label{fig:compare_models}
\end{figure*}

\section{Results}
\label{sec:results}

\subsection{Impact of He {\sc ii} absorption on the UVB}
\label{sec:helium_II_impact}
We begin in Figure \ref{fig:meanUVB} by comparing volume-averaged UVBs in models with and without \HeII at redshifts 10, 7, and 5. Figure~\ref{fig:meanUVB} shows only the galaxy contribution to the UVB even though the simulations do include flux from active galactic nuclei; the latter is included below. The Lyman-series absorption of \HeII has a significant impact on the mean UVB leading up to \HI reionization as the UVB absorbed by \HeII increases up to 3 dex in this range. The increased absorption is evident at $z=10$ and continues to strengthen as the \HeII fraction grows through z$\sim$5.

In Figure \ref{fig:ratioUVB}, we plot the ratio of the UVB intensity with and without \HeIIb. The linear panel accentuates the bump from the Lyman-$\alpha$ emission following absorption by a \HeII Lyman-series transition at 3 Ryd as well as the minor differences in intensity between redshifts at energies $>$4 Ryd. The logarithmic panel shows that \HeII reprocessing reduces the intensity ratio from z=10 to z=7 by 0.5 dex, and a total reduction of 3 dex from z=10 up to z=5.

\begin{figure}
    \includegraphics[width=0.49\textwidth]{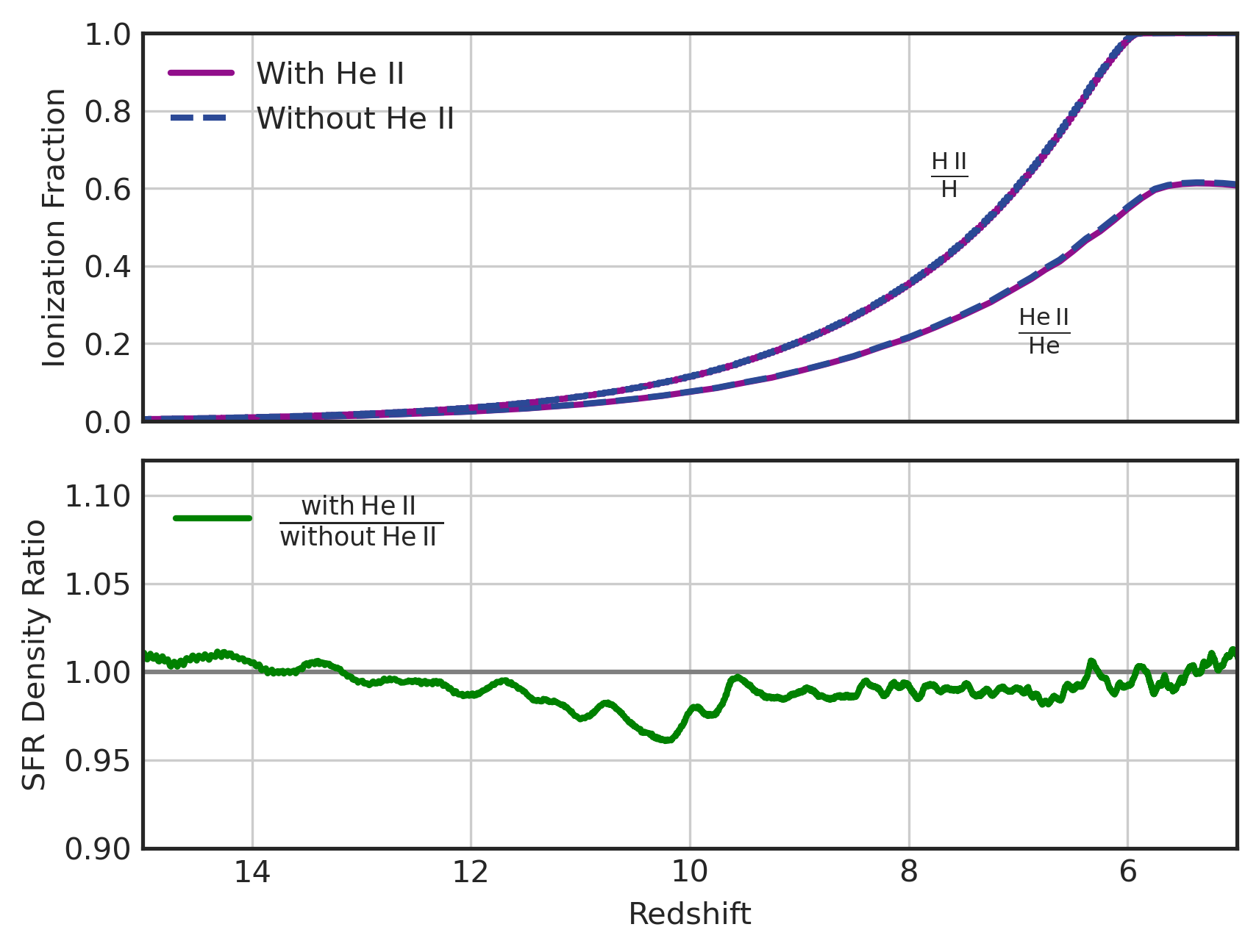}
    \caption{ (Top) The fraction of hydrogen that is H {\sc ii} and the fraction of helium that is He {\sc ii}, both with and without He {\sc ii} reprocessing. The addition of He {\sc ii} into the simulation alters the ionization fractions by $\sim1\%$. (Bottom) The ratio of star formation rate density between the two simulations showing a moderate impact on star formation.}
    \label{fig:fraction}
\end{figure}

\subsection{Overdensity}
\label{sec:overdensity}

The question of whether \HeII reprocessing impacts observable metal ions depends on how its strength varies with environment because metals inhabit overdense regions, which tend to have both higher gas density and emissivity. As a first step toward understanding the impact of \HeII reprocessing on metal absorbers, we therefore ask whether this interaction leads overdense regions to have stronger or weaker UVBs in the range affected by \HeII absorption and re-emission. In Figure \ref{fig:overdensity}, we divide the simulation volume into two groups of voxels via the ratio of density to the mean density: voids ($\Delta < 1$) and overdense regions ($\Delta > 5$) where $\Delta = \rho/\langle\rho\rangle$. In addition, we plot the UVB at the mean density for reference. 

At $z = 10$, overdense regions have a 2 dex stronger UVB because they host the majority of the star formation. This indicates that emissivity fluctuations are more important than opacity fluctuations. The more dense regions also show a weaker \HeII absorption feature at 3.5 Ryd as most of the helium present there has already further ionized to \HeIII. The inhomogeneity in the UVB vanishes at energies below 3 Ryd following the completion of \HI reionization ($z\approx6$). The IGM remains optically thick at higher energies, so less of this radiation reaches the voids and UVB inhomogeneity persists. We attribute this lingering high-energy inhomogeneity to the abundant \HeIIb-related opacity expected prior to the completion of \HeIIb~reionization at $z\sim3$, now apparent because three-dimensional simulations naturally treat its contributions from diffuse and dense gas realistically.

By contrast, underdense regions have a stronger \HeII absorption feature as the light in these voids has been more filtered in transit from the overdense regions. The inverse also remains true even at $z=5$: the \HeII absorption feature is weaker in overdensities because their flux has not been filtered as heavily. This shows one way that environment plays a role in the shape of the UVB.

Strong metal absorbers accompany the overdense gas near galaxies \citep{keating16} and they are sensitive to the spectral slope of the UVB. Models have shown that the number of high-ionization absorber systems, such as \CIVb, increases as the UVB become harder or more intense \citep{doughty18}. \citealt{finlator20} suggested, however, that if overdense regions are more opaque to higher energy UV, we may actually find more \CIV in underdense regions. As noted from Figure \ref{fig:overdensity}, we can see that the UVB is harder in overdense regions at all energies, in direct conflict with this speculation. Instead, most of the \CIV and other highly ionized metals will be found in overdense regions. It is also clear that the UVB range that ionizes \CIII into \CIV is still inhomogenous at $z = 5$, and we expect it to remain so until \HeII reionization completes at $z\approx3$.

\subsection{Comparison to 1D Models}

For context, we compare our three-dimensional, volume-averaged UVB with two popular one-dimensional models. In Figure \ref{fig:compare_models}, we show the mean UVB of \TD with and without \HeII alongside HM12 and FG20. We add the UV contribution from quasars to the galactic UV in our models to better compare to the 1D models which include both.

Both simulations show higher flux densities at energies below the \HeII ionization threshold (4 Ryd) prior to the completion of \HI reionization ($z\sim6$), but the significance of this discrepancy is difficult to assess given that TD is spatially-inhomogeneous. By $z=5$, all three models have converged at energies below 3.5 Ryd. Above 5 Ryd, all the flux comes from quasars so the difference is in the adopted quasar emissivity model.

A curious discrepancy arises at $z=5$ between 3.5--4 Ryd in Figure~\ref{fig:compare_models}. In this energy range, the UVB in the TD simulation that \emph{omits} \HeII opacity closely resembles predictions from HM12 and FG20, both of which \emph{include} it. Evidently, \HeII opacity is much stronger in our simulations. The discrepancy arises because HM12 and FG20 assume that the UV opacity is dominated by Lyman-limit systems at all times. While reasonable following the completion of \HeII reionization at $z\approx3$, this approach neglects the dominant contribution of more diffuse gas at earlier times. In the immediate aftermath of HI reionization, our simulations indicate that most He {\sc ii} resides in the diffuse gas between Lyman limit systems.

These results are consistent with~\citet{puchwein19}, who showed that the UVB is unaffected by diffuse gas following the completion of H {\sc i} and He {\sc ii} reionization, but the He {\sc ii} absorption feature is much stronger at $z=3.6$ when the diffuse gas is accounted for (see their Figure 5).

\begin{figure}
    \includegraphics[width=0.49\textwidth]{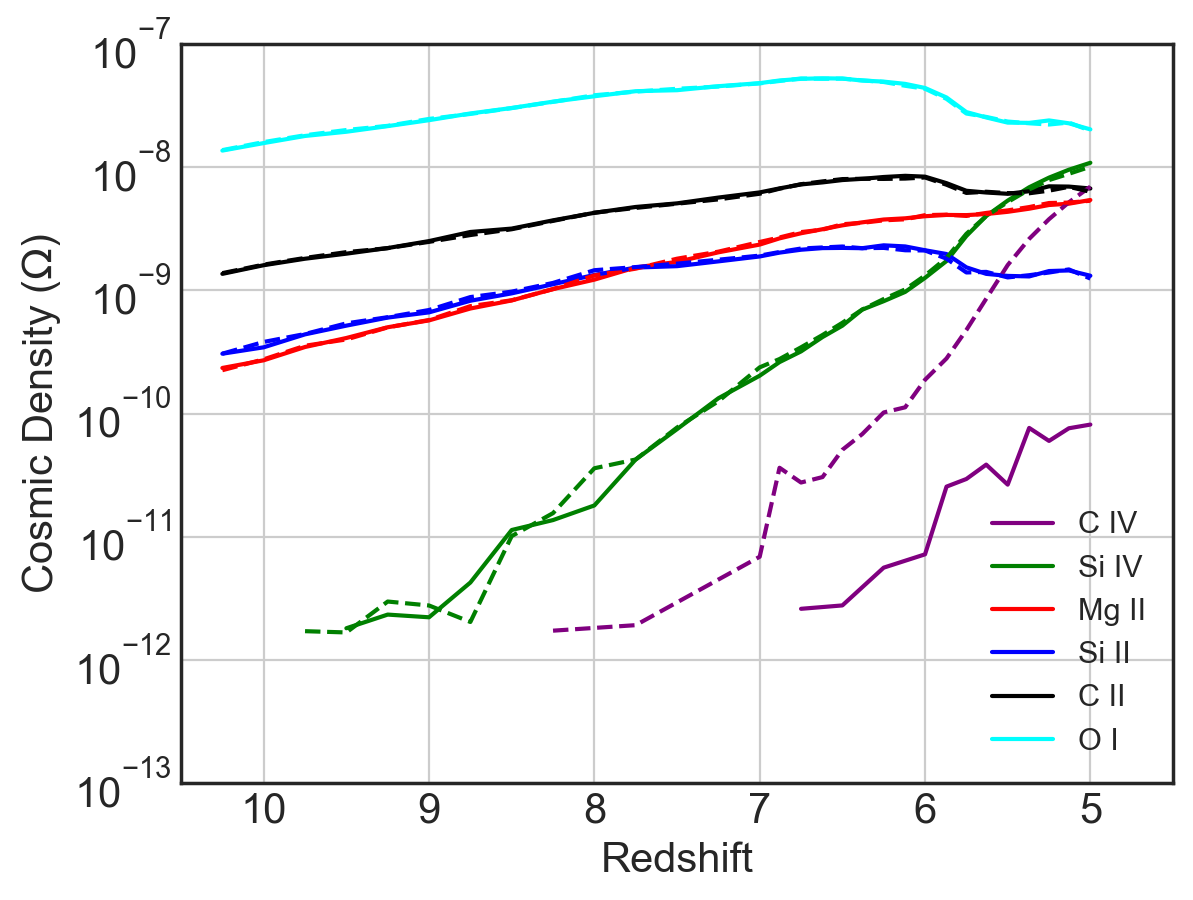}
    \caption{ The cosmic density of 6 different ions at z=10 to z=5. Solid lines are simulations that include \HeII reprocessing while dashed lines do not. \CIV density drops significantly due to \HeIIb, especially at lower redshifts. \SiIV shows minor differences up to z=7.5 and matches closely at lower redshifts. The change in \MgII, \SiII, \CII~ and \OI ions are all negligible.}
    \label{fig:cosmic}
\end{figure}

\subsection{Star Formation} \label{sec:sfd}

The cosmic star formation rate density is sensitive to \HeII reprocessing because decreases in UVB intensity from \HeII absorption suppress IGM heating, promoting new star formation in low-mass halos. In Figure \ref{fig:fraction}, we quantify this by showing that the ratio of the star formation rate densities varies up to 5\% in the model with \HeII to the one without \HeIIb. The decrease at $z\sim10$ may be a product of increased ionization in the model without \HeIIb: the larger electron fraction promotes collisional excitation cooling, which in turn boosts gas accretion and star formation, particularly in low-mass halos. The increased star formation, in turn, accelerates reionization. This positive feedback effect is quite weak, however, with the \HeII model increasing the overall star formation rate density and the \HeII reionization rate by $\sim1\%$ after $z=5$.

\subsection{Impact on Metal Ions in the CGM} \label{sec:ions}

 Metals are more sensitive to high-energy UV than neutral hydrogen, and the additional opacity from He {\sc ii} is likelly to reduce the abundances of ions in higher ionization states \citep{mh09}. We now explore this to see how much changes in the UVB associated with He {\sc ii} reprocessing affect the resulting abundances of metal absorbers, which are in turn extracted from our simulations as described in Section~\ref{ssec:rayCasting}.
 
In Figure \ref{fig:cosmic}, we compare the cosmic ion mass densities ($\Omega$) of 6 ionic species from 10 < z < 5 normalized to the critical mass density with models including \HeII reprocessing (solid lines) and without it (dashed lines). A minimum column density cutoff is implemented of $10^{13.2}$ cm$^{-2}$ for \CIV, $10^{12.5}$ cm$^{-2}$ for \SiIV, and $10^{13}$ cm$^{-2}$ for other ions. There is no change in $\Omega_{{\rm Mg II}}$, $\Omega_{{\rm Si II}}$, $\Omega_{{\rm C II}}$ and $\Omega_{{\rm O I}}$, in agreement with previous results that low-ionization ions are relatively insensitive to the details of feedback and ionization~\citep{keating16}. There may be a minor impact to $\Omega_{{\rm Si IV}}$ at $z \ge 8$, but this is difficult to assess as Si IV absorbers are rare at such redshifts. The \CIV density exhibits the only considerable change, differing by 1 dex at $z\sim7$, and nearly 2 dex at $z\sim5$ between the two simulations.

\begin{figure}
    \includegraphics[width=0.49\textwidth]{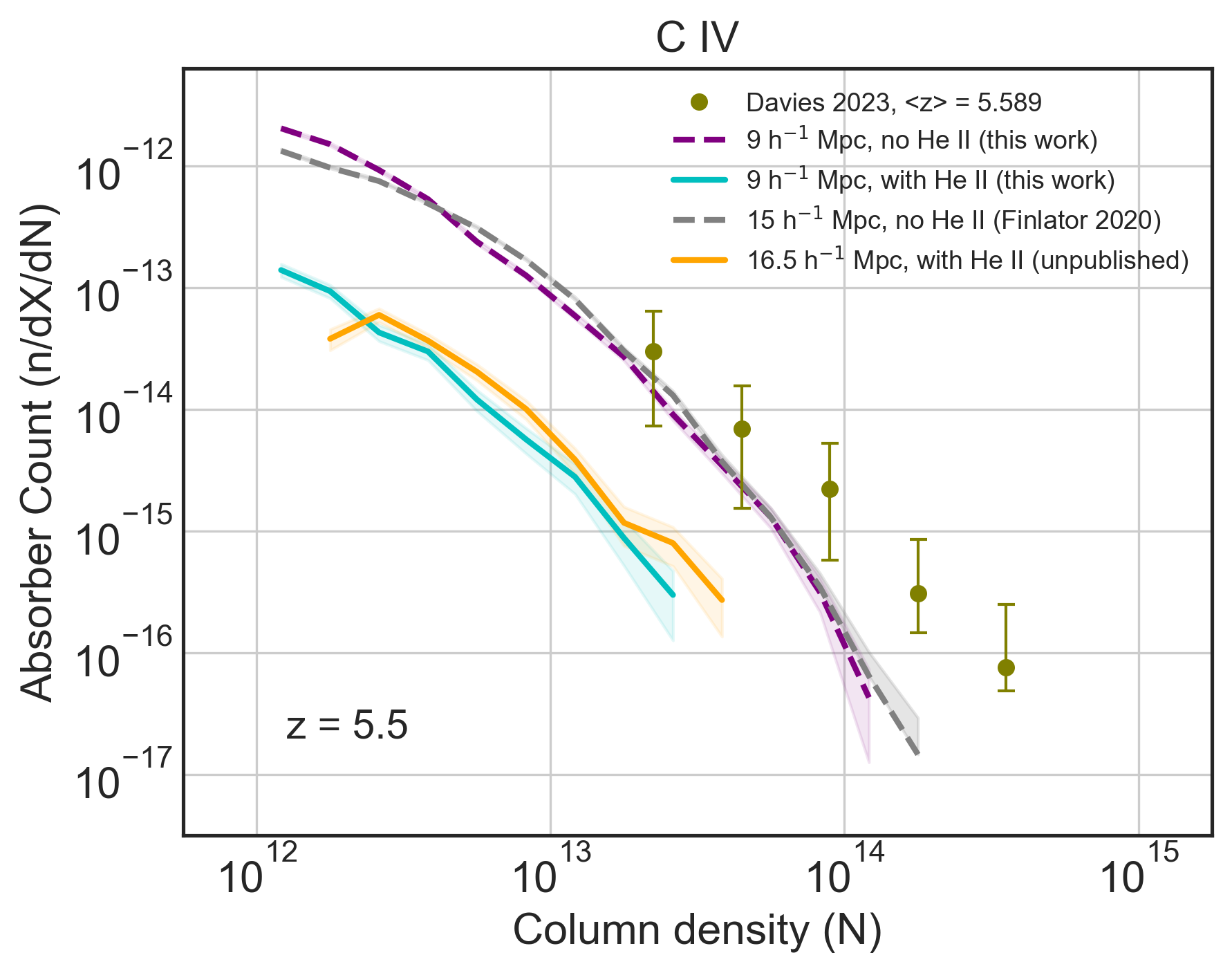}
    \caption{ The column density of \CIV absorbers from the two TD simulation runs in this work, the original TD run analyzed in \citealt{finlator20}, and a newer unpublished TD run in a larger 16 Mpc volume. One sigma errors included as $\sqrt N$. The addition of \HeII reduces the absorber count by $\sim$1 dex at all column densities available for comparison, widening the discrepancy between the models and the observations of \citealt{davies23}. }
    \label{fig:civ}
\end{figure}
\begin{figure}
    \includegraphics[width=0.49\textwidth]{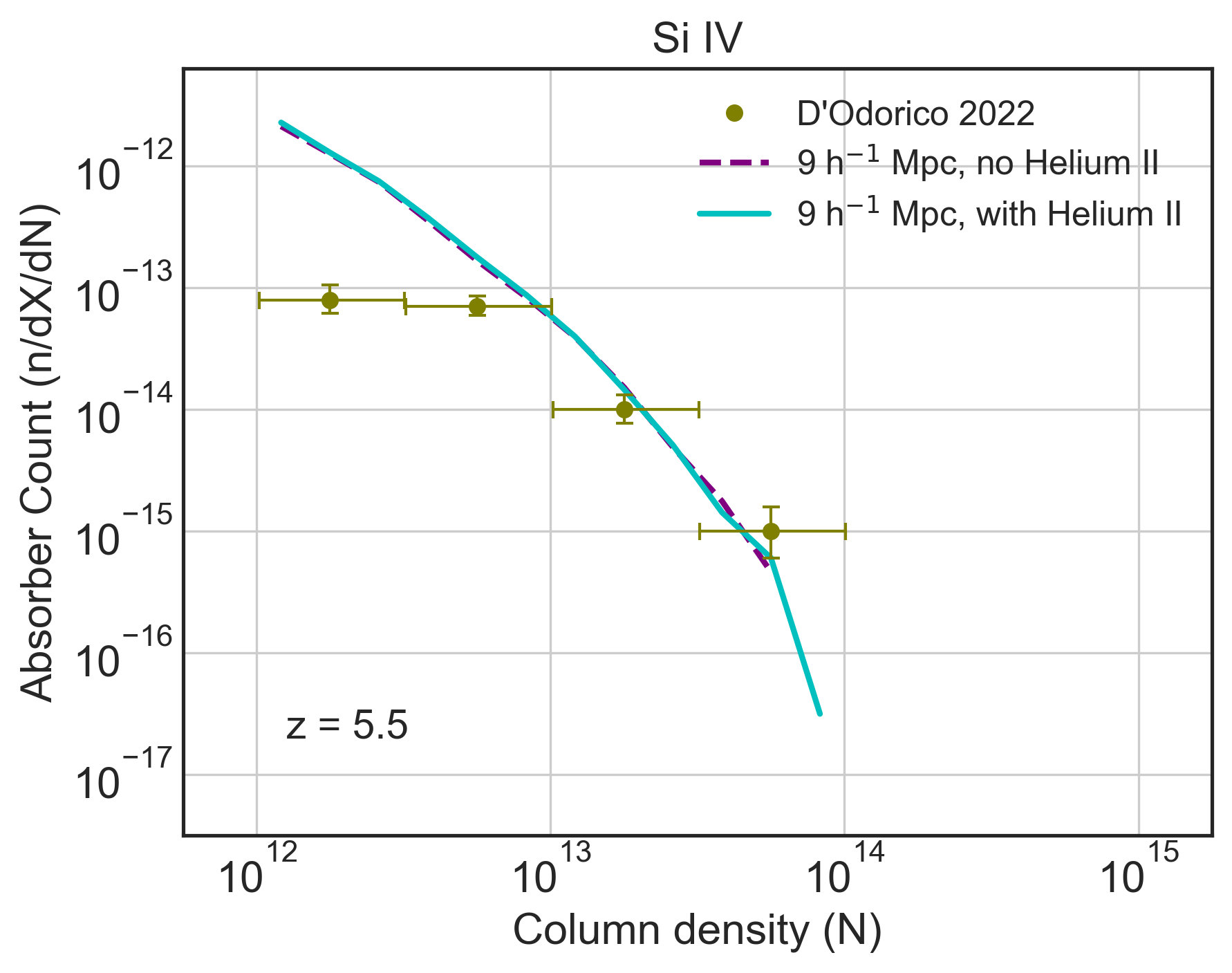}
    \caption{ The column density of \SiIV absorbers. Turning on the He {\sc ii} absorption had negligible impact in the TD simulations. Weak absorbers are overproduced by up to 1 dex in the simulations, but stronger absorbers ($>10^{13}$) show excellent agreement to the observational data from \citealt{dodorico22}. }
    \label{fig:siiv}
\end{figure}

In Figures \ref{fig:civ} and \ref{fig:siiv}, we quantify the impact of \HeII re-processing on the circumgalactic medium (CGM) via the \CIV and \SiIV absorber column distributions (CDDs), using the same TD models with and without \HeII reprocessing. Figure \ref{fig:civ} additionally compares our simulations to a previous 15 $h^{-1}$ Mpc run~\citep{finlator20}, which does not treat \HeII re-processing, as well as a newer 16.5 $h^{-1}$ Mpc simulation that is identical to our \HeII 9 $h^{-1}$ simulation but subtends a larger volume at the same mass resolution. The addition of \HeII reduces the absorber count by $\sim$1 dex at all column densities available for comparison. There is good agreement between the smaller and larger volumes, indicating that box size limitations do not affect the CDD for $N<10^{14}$cm$^{-2}$. The 9 and 15 $h^{-1}$ Mpc runs without \HeII are likewise in good agreement, though the newer 9 $h^{-1}$ Mpc run may show a slightly elevated production of weak absorbers~\citep[see also][]{doughty21}. The underproduction of stronger \CIV absorbers, previously noted by \citealt{finlator20}, becomes an even greater discrepancy with \HeII reprocessing included. This persistent problem is worth increased theoretical attention.

In Figure \ref{fig:siiv}, we show that, in contrast to CIV, the \SiIV CDD shows minimal impact from the addition of \HeII opacity, although weak absorbers remain marginally overproduced compared to observations from \citealt{dodorico22}. The abundance of strong absorbers ($N > 10^{13}$cm$^{-2}$) is well-reproduced. The insensitivity of \SiIVb~to \HeII reprocessing while \CIV shows a considerable change confirms that \SiIV is not affected by UVB inhomogeneity following the completion of HI reionization while \CIV is \citep{finlator2016}. This is reasonable to expect during the epoch when the mean free path to HI-ionizing flux is long whereas that of \HeII-ionizing flux remains short. It has been shown previously~\citep{doughty18, finlator20} that a similar model which produces enough stars, a realistic UVB, and a reasonable reionization history continually underproduces \CIV but matches \SiIVb, just as we see here. We conclude that the interaction of \HeII and the UVB only decreases \CIV abundance, exacerbating the \CIV underproduction problem in cosmological simulations.

\section{Summary and Conclusions}  \label{sec:summary}

\vspace{3mm}

Our key findings are:  

\begin{enumerate}{}{\leftmargin=1em}

     \item The suggestion by HM12 that \HeII opacity has a significant impact on \CIV abundance is confirmed. Si {\sc iv}, C {\sc ii}, Mg {\sc ii}, Si {\sc ii}, and O {\sc i} are not impacted significantly.

     \item He {\sc ii} reprocessing reduces the volume-averaged UVB intensity between 3.5 and 4 Ryd by 30$\times$ at $z \sim 7$. By $z \sim 5$, this reduction reaches a maximum of 1000$\times$.
    
     \item Prior to HI reionization, overdense regions have up to 100$\times$ the average UVB intensity from 1--10 Ryd. Following HI reionization ($z<6$), the difference is seen only at $>$4 Ryd and is reduced to $\sim$10$\times$. Consequently, \CIV should trace predominantly overdense regions at early times.

     \item At $z=5$ in the post-\HI reionization Universe, there is excellent agreement with the HM12 and FG20 models at < 3 Ryd, reflecting agreement in the modeled \HI reionization history. However, our simulations predict a dramatic drop at 3.5--4 Ryd from \HeII reprocessing and then consistently lower UVB intensity > 4 Ryd where TD accounts for both overdense and underdense gas.
    
\end{enumerate}


\vspace{3mm}


Modeling reionization in three dimensions as a spatially inhomogeneous process allows simulations like \TD to maintain a connection between the ionizing radiation and the star-forming galaxies that produced it which provides insight into the properties of the IGM during this important period. The key problem is capturing the separately-evolving contributions of diffuse and moderately-overdense gas to the overall opacity. Existing UVB synthesis models which assume \HeII exists only in Lyman-limit systems likely underestimate its impact by neglecting the diffuse \HeII component and should incorporate \HeII reprocessing as done by \citealt{puchwein19} to maintain realistic \HeII opacity. Where possible, spatial inhomogeneity in the UVB should be emphasized in future research because fluctuations impact the abundance of \HeII which in turn impacts \CIVb, a key observable in the CGM.

Our predicted abundances of strong \CIV absorbers are much lower than the latest observational results from \citealt{davies23}. \CIV has proven to be highly sensitive to many factors such as the treatment of the escape fraction which is further complicated by limits in a simulation's dynamic range. It seems that overall changes to metal yields can be ruled out by the reasonable agreement with observed \SiIVb. Our work points to the need either for increased carbon yields or increased generation or escape of high-energy flux. We now consider a number of possibilities to address this discrepancy.

First, it has been speculated that shocks from supernovae may impact the CGM and create more high-energy photons for the UVB \citep{li20}. This could be a consequential factor in localized \CIV abundances that are actively observed in CGM absorption lines. To our knowledge, this has not been modeled in large-scale simulations.

Second, JWST's surprising discovery of high-redshift compact galaxies containing massive dust-obscured AGN, dubbed ``little red dots", may provide additional ionizing flux not currently accounted for \citep{Harikane23}. Early data releases suggest they may have been plentiful during the epoch of reionization \citep{Kokorev24a} and future modeling can incorporate a treatment for these interesting UV sources.

Third, a treatment of the escape fraction that is density-bounded can harden the UVB and allows galaxies to ionize their surroundings more efficiently. This can produce additional \CIV in the CGM without affecting \SiIV abundances \citep{finlator20}. It is likely that high-z star-forming galaxies will be surrounded by inhomogenously ionized gas \citep{Nakajima14} and a combination of ionization-bounded and density-bounded photon escape is probably realistic.

Lastly, improved modeling of AGB stars may reveal additional carbon than our simulations currently reflect which may impact the abundances of the carbon ions. Because the AGB stage of stellar evolution is short-lived, observations for comparing to models are difficult to obtain and therefore modeling mass loss rates is not well-constrained.

\TD is producing an accurate reionization history, reasonable IGM densities and temperatures, and expected star formation rate densities, but the inaccuracy of the \CIV production remains an open question. Addressing the \CIV anomaly should be emphasized in future research as this ion could prove to be an excellent probe of \HeII reionization, particularly because \HeII is not directly observable from the ground at redshifts above $\sim3$.


\subsection{Future Work}
While the impact of the He {\sc ii} absorption is evident in Figure \ref{fig:meanUVB}, we note that the sawtooth features as shown by \citep{mh09} are not readily apparent, as \TD's 32 UV bins currently allow for just 4 within the range of 3-4 Ryd. We expect that future runs at higher spectral resolution will more clearly show the predicted ``sawtooth" structure.

Recombinations from free-bound transitions in \HeII can result in Lyman-$\alpha$ emissions (process iv in Section~\ref{sec:intro}). This process, not yet accounted for in \TD, would have no impact on \CIVb, but it could have a modest impact on \SiIVb. Hence while its absence does not impact our current conclusions, we plan to fold it in.

\vspace{5mm}

\section{Acknowledgements}

\textbf{EH}: Conceptualization, Software, Investigation, Visualization, Writing - Original draft; \textbf{KF}: Conceptualization, Supervision, Software, Writing - Review \& Editing, Project administration, Funding acquisition; \textbf{SK}: Writing - Review \& Editing; Visualization; \textbf{MS}: Writing - Review \& Editing.

EH would like to thank Rebecca Davies for generously providing her most up-to-date \CIV abundances even before they were published. This work utilized resources from the New Mexico State University High Performance Computing Group, which is directly supported by the National Science Foundation (OAC-2019000), the Student Technology Advisory Committee, and New Mexico State University and benefits from inclusion in various grants (DoD ARO-W911NF1810454; NSF EPSCoR OIA-1757207; Partnership for the Advancement of Cancer Research, supported in part by NCI grants U54 CA132383 (NMSU)). KF and EH gratefully acknowledge support from STScI Program \#HST-AR-16125.001-A. Support for this program was provided by NASA through a grant from the Space Telescope Science Institute, which is operated by the Associations of Universities for Research in Astronomy, Incorporated, under NASA contract NAS5-26555. The Cosmic Dawn Center is funded by the Danish National Research Foundation.

\section{Data Availability}

These results were generated from the \TD cosmological simulation suite. Snapshots with particle and UVB information and log files with summary statistics are available upon email request.

\bibliographystyle{mnras}
\bibliography{HeliumII_Final.bib} 

\end{document}